\documentclass[aps,prl,reprint]{revtex4-1}
\usepackage{times}
\usepackage{graphicx}
\usepackage{dcolumn}
\usepackage{bm}
\usepackage{amsmath}
\usepackage{amsfonts}
\usepackage{setspace}
\usepackage{braket}
\usepackage{xcolor}
\usepackage{soul}
\providecommand{\U}[1]{\protect\rule{.1in}{.1in}}
\begin{document}

\preprint{APS/123-QED}

\title{Non-Hermitian Quantum Fermi Accelerator}
\author{Andreas Fring$^a$ and Takano Taira$^b$}

\affiliation{$^a$Department of Mathematics, City, University of London,\\
Northampton Square, London EC1V 0HB, UK \\
$^b$Research Fellow of Japan Society for the Promotion of Science,\\
Institute of Industrial Science, The University of Tokyo,\\
5-1-5 Kashiwanoha, Kashiwa 277-8574, Japan.\\
E-mail: a.fring@city.ac.uk, taira904@iis.u-tokyo.ac.jp
}
\date{\today}
\begin{abstract}
    We exactly solve a quantum Fermi accelerator model consisting of a time-independent non-Hermitian Hamiltonian with time-dependent Dirichlet boundary conditions. A Hilbert space for such systems can be defined in two equivalent ways, either by first constructing a time-independent Dyson map and subsequently unitarily mapping to fixed boundary conditions or by first unitarily mapping to fixed boundary conditions followed by the construction of a time-dependent Dyson map. In turn this allows to construct time-dependent metric operators from a time-independent metric and two time-dependent unitary maps that freeze the moving boundaries. From the time-dependent energy spectrum, we find the known possibility of oscillatory behavior in the average energy in the PT-regime, whereas in the spontaneously broken PT-regime we observe the new feature of a one-time depletion of the energy. We show that the PT broken regime is mended with moving boundary, equivalently to mending it with a time-dependent Dyson map.
\end{abstract}

\maketitle
\section{Introduction}
    Classical versions of Fermi accelerators were originally proposed by Fermi \cite{Fermi} more than seventy years ago as a possible explanation for the high energies observed in cosmic radiation. The simplest classical Fermi accelerator model consists of a free particle moving between two walls simulating magnetic fields, with one of them fixed and the other moving in time, with the collisions between the particle and the walls being perfectly elastic. Besides predicting features of cosmic rays in the spirit of the original motivation, such as the maximum energy that particles can reach is proportional to the strength of the magnetic field and the size of the acceleration region, the models were also found to exhibit classical chaotic behavior \cite{ulam61,zaslav72,lichtenberg80}. The latter is due to the fact that the description in phase-space of consecutive scatterings between the walls and the particle leads to nonlinear maps, which in their simplest version, corresponding to the so-called Ulam maps. For a recent overview of the latest experimental observations of ultra-high energy cosmic rays, see for instance \cite{gora2018pierre}.

    Quantum versions of Fermi accelerator models are set up in a similar fashion, described by the Schr\"odinger equation with Dirichlet boundary conditions. They allow us to study quantum chaos \cite{karner89,seba90} and other interesting phenomena \cite{doescher1969,munier1981,pinder1990,makowski1991,pereshogin1993,dodonov1993}, such as the possibility of an energy gain in the time-dependent spectrum. Here the purpose is to investigate such a system with the starting Hamiltonian taken to be non-Hermitian, but $\mathcal{PT}$-symmetric/pseudo-Hermitian. In the broken $\mathcal{PT}$-regime we observe {the previously unseen non-periodic nature of the average energy over time}.    

    Our starting point is to consider a time-independent $\mathcal{PT}$-symmetric/pseudo-Hermitian \cite{bender1998real,mostafazadeh2002pseudo} Hamiltonian $\Tilde{H} = -\frac{\hbar^2}{2m}\partial_x^2 + \Tilde{V}(x)$, where $\Tilde{V}(x)$ is a non-Hermitian potential. The Schrödinger equation with Dirichlet boundary condition is given by
    \begin{equation}
        i\hbar\frac{\partial}{\partial t}\Tilde{\psi}(t,x) = \Tilde{H}(x)\Tilde{\psi}(t,x), \qquad \Tilde{\psi}(t,\pm \ell) = 0,\label{Eq: Time-dependent Schrodinger equation}
    \end{equation}
    where $\ell >0$.
    The Hilbert space of the system consists of square-integrable functions in the interval $[-\ell,\ell]$, i.e., $\Tilde{\psi}(t,x) \in L^2([-\ell,\ell])$. This Hamiltonian is said to be $\mathcal{PT}$-symmetric if the Hamiltonian and the wave functions are symmetric under an anti-linear transformation, such as $p\rightarrow p$, $x\rightarrow -x$, and $i \rightarrow -i$, in our case.
    
    The standard procedure in $\mathcal{PT}$-symmetric quantum mechanics is to map the non-Hermitian Hamiltonian (\ref{Eq: Time-dependent Schrodinger equation}) to a Hermitian Hamiltonian with a Dyson map $\eta$ such that $\Tilde{H} = \eta \Tilde{h}\Tilde{\eta}^{-1} \not= \Tilde{H}^\dagger$, $\Tilde{h}^\dagger = \Tilde{h}$. {Recall that the Dyson map $\eta$ is generally non-unique. Extensive discussion on the uniqueness of $\eta$ can already be found in \cite{scholtz1992quasi}, where the authors demonstrated that $\eta$ is uniquely fixed by demanding the irreducibility of some set of operators. In the case of the Swanson model \cite{swanson2004transition}, it was equivalently shown in \cite{musumbu2006choice} that the uniqueness of  $\eta$ can be ensured by requiring the Hamiltonian and one other operator (e.g. position, momentum, or number operator) to correspond to their Hermitian counterpart.}
    
    Let us denote the new wave function $\Tilde{\phi}(t,x) = \Tilde{\eta} \Tilde{\psi}(t,x)$ where the non-Hermitian operator $\Tilde{\eta}$ is time-independent. Therefore the Schr\"{o}dinger equation and the boundary condition (\ref{Eq: Time-dependent Schrodinger equation}) are simply mapped to 
    \begin{eqnarray}
        i \hbar \frac{d}{dt} \Tilde{\phi} (t,x) = \Tilde{h}(x)\Tilde{\phi}(t,x),\qquad \Tilde{\phi}(t,\pm \ell)=0.\label{Eq: Time-independent Hermitian Schrodinger equation}
    \end{eqnarray}
    In $\mathcal{PT}$-symmetric quantum mechanics inner product in the Hilbert space needs to be redefined. Accordingly, the average energy of the non-Hermitian Hamiltonian is given by $\braket{E}_\eta =: \int_{-\ell}^{\ell} dx \Tilde{\psi} \tilde{\rho} \tilde{H} \Tilde{\psi}$, where the Hermitian positive definite metric is defined as $\Tilde{\rho} :=\Tilde{\eta}^\dagger \tilde{\eta}$. This can be rewritten in terms of the Hermitian Hamiltonian as 
    \begin{eqnarray}\label{Eq. average energy}
        \braket{E (t)}_{\eta} :=  \int_{-\ell}^{\ell} dx \Tilde{\psi}^\dagger\Tilde{\rho}   \Tilde{H}\Tilde{\psi} =  \int_{-\ell}^{\ell} dx \Tilde{\phi}^\dagger \Tilde{h}\Tilde{\phi}.
    \end{eqnarray}
    The common characteristic of the non-Hermitian system is that the above equality only holds when the non-Hermitian Hamiltonian and the wave function are $\mathcal{PT}$-symmetric. The average energy of $\Tilde{H}$ acquires complex conjugate eigenvalues in the $\mathcal{PT}$-broken regime. {However, we will show that when the boundary $\ell$ is time-dependent, the average energy is defined above square real energy, even in the $\mathcal{PT}$-broken regime.}

    It has been established that real-valued average energies can be obtained in all regimes when the non-Hermitian Hamiltonian or the Dyson map are time-dependent \cite{fring2017mending}. {See also the review of the time-dependent non-Hermitian quantum mechanics \cite{fring2023int}.} In this work, we demonstrate that real-valued average energies can also be attained in the $\mathcal{PT}$-broken regime of the Swanson model by introducing time-dependence to the boundary condition $\ell$, instead of the Hamiltonian or Dyson map. Moreover, we establish the equivalence of our approach with a previous method \cite{fring2017mending}, where the Dyson map's time-dependence was used to mend the $\mathcal{PT}$-broken regime. Our two primary findings are summarized in Table \ref{tab:my_label} and Fig. \ref{fig: The square}. We will provide the explicit derivation of the scheme in Fig. \ref{fig: The square} in the next section.
        \begin{table}[]
            \centering
            \begin{tabular}{c|c|c}
                 & $\mathcal{PT}$-symmetric&$\mathcal{PT}$-broken\\\hline
                Time-independent boundary & $\braket{E}_\eta \in \mathbb{R}$ &$\braket{E}_\eta \in \mathbb{C}$ \\\hline
                 Time-dependent boundary & $\braket{E}_\eta \in \mathbb{R}$ &$\braket{E}_\eta \in \mathbb{R}$\\
            \end{tabular}
            \caption{The average energy $
            \braket{E}_\eta$ defined in Eq. (\ref{Eq. average energy}) is compared for two $\mathcal{PT}$-regimes with time-dependent/independent boundary conditions. In both cases there is a phase transition in the dynamical behavior of $\braket{E}_\eta$ between $\mathcal{PT}$ broken/unbroken regimes. The detail is presented in section \ref{section: Swanson model}.}
            \label{tab:my_label}
        \end{table}
\section{Equivalence of time-dependent boundary and Dyson map}
    Let us assume that the boundary is time-dependent, i.e., $\ell = \ell(t)$, then, the wave functions $\Tilde{\psi}(t,x)$ and $\Tilde{\psi}(t',x)$ belong to two different Hilbert spaces for $t\not= t'$. Therefore, the time derivative of the wave function does not belong to any Hilbert space for any time slice, which implies that the above Schrödinger equation is not well-defined. However, in \cite{di2013quantum}, the problem was resolved by formally embedding the system into a larger domain $L^2 (\mathbb{R}) =L^2([-\ell,\ell])\oplus L^2((-\infty, -\ell)\cup (\ell,\infty))$, where extended Hamiltonian is $\tilde{H} (x) \oplus 0$. This embedding implies that the integration contour of the average energy (\ref{Eq. average energy}) can be understood as 
    \begin{eqnarray}
        \braket{E(t)}_\eta &=& \int_{-\infty}^{\infty} dx \Tilde{\psi}^\dagger \Tilde{\rho} \left[\tilde{H} (x) \oplus 0\right] \Tilde{\psi}\nonumber\\
        &=&  \int_{-\ell(t)}^{\ell(t)} dx \Tilde{\psi}^\dagger \Tilde{\rho} \left[\tilde{H} (x)\right] \Tilde{\psi} .
    \end{eqnarray}
    
    To remove the time dependence of the boundary from the Hilbert space, a time-dependent unitary operator $U(t)$ is introduced as
    \begin{equation}\label{Eq. Unitary map 1}
        \begin{array}{cccc}
            U: &L^2 (\mathbb{R})& \rightarrow&L^2 (\mathbb{R})\\
             &f(t,x) &\rightarrow & \quad\sqrt{\ell (t)} f\left(t,\ell(t) x\right),
        \end{array}
    \end{equation}
    which maps all wave functions in $L^2 (\mathbb{R})$ to $L^2 (\mathbb{R}) = L^2([-1,1])\oplus L^2((-\infty, -1)\cup (1,\infty))$, thereby removing the time dependence of the boundary from the Hilbert space. The factor $\sqrt{l(t)}$ is necessary to ensure the transformation is unitary. The Hamiltonian is mapped to $U\Tilde{H}U^\dagger  \oplus 0$. For the rest of the paper, we will drop the $0$ component of the extended operators for brevity.

    Let us define the unitary transformed wave function as $U(t)\Tilde{\psi} (t,x) =: \psi(t,x)$. The time-dependent Schr\"{o}dinger equation (\ref{Eq: Time-dependent Schrodinger equation}) is also mapped by the unitary operator as 
    \begin{eqnarray}
        i\hbar\frac{\partial}{\partial t}\psi(t,y) &=& \left(U\Tilde{H}(x)U^\dagger +i \hbar U \partial_t U^\dagger\right)\psi(t,y)\nonumber\\
        &=&  \left[\Tilde{H}(y) + \frac{\partial_t \ell}{2\ell} \left\{y,i \hbar \partial_y\right\}\right] \psi(t,y)\nonumber\\
        &=: &H(t,y) \psi(t,y) \label{Eq: Time-dependent non-Hermitian Hamiltonian}
    \end{eqnarray}
    where $y = \ell x \in [-1,1]$.

    Alternatively, assuming pseudo-Hermiticity, the time-independent non-Hermitian Hamiltonian (\ref{Eq: Time-dependent Schrodinger equation}) can be mapped to a Hermitian Hamiltonian via Dyson map $\Tilde{H} = \eta \Tilde{h}\Tilde{\eta}^{-1}$, $\Tilde{h}^\dagger = \Tilde{h}$. {Note that when we consider the $\mathcal{PT}$-broken regime, the pseudo-Hermicity is broken and the Hamiltonian $\tilde{h}$ becomes non-Hermitian. However, we will show that even in such a case, the average energy (\ref{Eq. average energy}) is real due to the time-dependent boundary}. Let us denote the new wave function $\Tilde{\phi}(t,x) := \Tilde{\eta} \Tilde{\psi}(t,x)$ where the non-Hermitian operator $\Tilde{\eta}$ is time-independent. Therefore the Schr\"{o}dinger equation and the boundary condition (\ref{Eq: Time-dependent Schrodinger equation}) are simply mapped to 
    \begin{eqnarray}
        i \hbar \frac{d}{dt} \Tilde{\phi} (t,x) = \Tilde{h}(x)\Tilde{\phi}(t,x),\qquad \Tilde{\phi}(t,\pm \ell)=0.\label{Eq: Time-dependent Hermitian Schrodinger equation}
    \end{eqnarray}
    Then the above procedure to remove the boundary time dependence can be applied to the mapped Hermitian system, and one obtains 
    \begin{eqnarray}
        i\hbar \frac{d}{dt} \phi(t,y)&=& \left(u \Tilde{h} u^\dagger - i \hbar u \partial_t u^{\dagger}\right)\phi(t,y)
        \nonumber\\
        &=&\left[\Tilde{h}(y) + \frac{\partial_t \ell}{2\ell} \left\{y,i \hbar \partial_y\right\}\right]\phi(t,y)\nonumber\\
        &=& h(t) \phi(t,y) ,\label{Eq: Time-dependent Hermitian Hamiltonian}
    \end{eqnarray}
    where $u \Tilde{\phi}(t,x)  = \phi(t,y)$ and $u$ is also defined in a same way as Eq. (\ref{Eq. Unitary map 1}). 

\begin{figure*}[t]
        \centering
        \begin{minipage}[b]{0.8\textwidth}           \includegraphics[width=\textwidth]{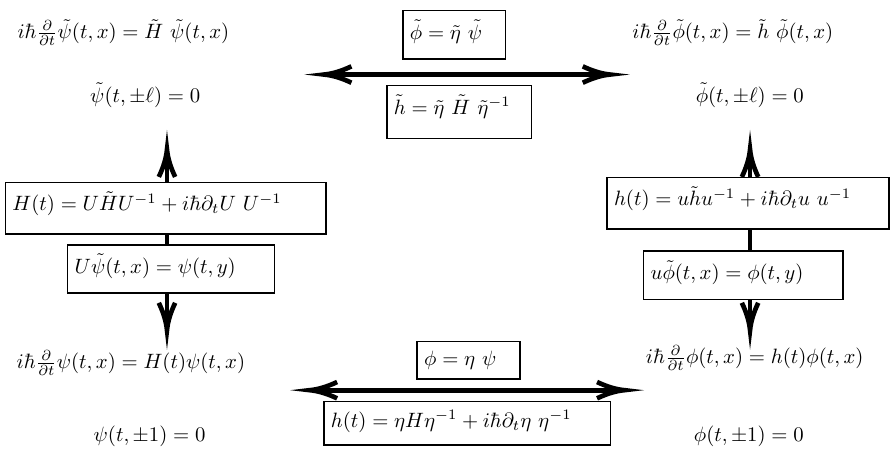}
        \end{minipage}
        \caption{Commutative scheme, showing the relations between two time-independent Schrödinger equations with time-dependent boundary conditions on the top row, and two time-dependent Schrödinger equations with time-independent boundary conditions on the bottom row.}
        \label{fig: The square}
    \end{figure*}

    It has been shown that in the time-dependent case {\cite{fring2016non}}, the Dyson map between non-Hermitian and Hermitian operators is given by 
    \begin{eqnarray}
        H(t) := \eta(t) h(t)\eta(t)^{-1} + i \hbar \frac{\partial \eta}{\partial t} \eta^{-1},
    \end{eqnarray}
    where the Dyson map $\eta(t)$ is a time-dependent non-Hermitian operator. We summarise the relation between Eqs.(\ref{Eq: Time-dependent Schrodinger equation}), (\ref{Eq: Time-dependent non-Hermitian Hamiltonian}), (\ref{Eq: Time-dependent Hermitian Schrodinger equation}) and (\ref{Eq: Time-dependent Hermitian Hamiltonian}) in the Fig. \ref{fig: The square}. Note that by requiring the scheme in Fig. \ref{fig: The square} to be commutative, we find the relation between two similarity transformations and the two Dyson maps
    \begin{eqnarray}\label{eq: relation between two similarity transformations}
        \eta(t) = u(t) \Tilde{\eta} U^\dagger (t),
    \end{eqnarray}
    which leads to the equivalence between the non-Hermitian time-dependent boundary problem (\ref{Eq: Time-dependent Schrodinger equation}) and the non-Hermitian time-dependent Hamiltonian problem with a time-dependent metric (\ref{Eq: Time-dependent non-Hermitian Hamiltonian}) discussed in \cite{fring2017mending}.
    
    Once we obtain the time-dependent Hermitian Hamiltonian, the average energy (\ref{Eq. average energy}) can be calculated. Using the relations in Fig. \ref{fig: The square} and the Eq. (\ref{eq: relation between two similarity transformations}), one can write down four alternative formulations of the average energy (\ref{Eq. average energy}). 
    \begin{eqnarray}
        \braket{E(t)} &=&\int_{-\ell(t)}^{\ell(t)}dx \tilde{\psi}^\dagger \tilde{\rho} \Tilde{H}\Tilde{\psi}=\int_{-\ell(t)}^{\ell(t)}dx \Tilde{\phi}^\dagger \Tilde{h}\Tilde{\phi}\label{Eq. Average energy 1}\\
        &=& \int_{-1}^{1} dx \phi^\dagger \left(h-i \hbar u_t u^\dagger\right)\phi\label{Eq. Average energy 3}\\
        &=& \int_{-1}^1 dx \psi^\dagger \eta^\dagger \eta \left[H + i \hbar \left(u^\dagger \eta\right)^{-1} \partial_t \left(u^\dagger \eta \right)\right]\psi .\qquad\label{Eq. Average energy 4}
    \end{eqnarray}
    {The operator inside the square brackets in Eq. (\ref{Eq. Average energy 4}) is called the energy operator. It was initially introduced in Ref. \cite{fring2016non}, serving as an isospectral operator in relation to the Hermitian operator procured through the time-dependent Dyson mapping of a non-Hermitian Hamiltonian. The implementation of this operator addresses the non-isospectral characteristic of the non-Hermitian Hamiltonian and its Hermitian counterpart, a discrepancy that arises due to the time dependence of the Dyson map.}
    
    We will apply these {general} relations to {a specific example that we choose to be} the Swanson model in the next section. 

\section{Swanson model: Mending $\mathcal{PT}$-broken regime via moving boundary}\label{section: Swanson model}
    {Typically, finding the exact Dyson map poses a substantial challenge, given that it necessitates solving an operator-valued algebraic equation. The Swanson model \cite{swanson2004transition} is one of the rare cases wherein multiple metrics have been found \cite{musumbu2006choice}, even in the time-dependent case \cite{fring2016non}. Exploiting this characteristic, we will compute the average energy corresponding to three distinct metrics, showing the energy spectrum is, indeed, real in all three instances.}
    
    {Furthermore, we will show that} a time-dependent boundary can lead to real average energy in both $\mathcal{PT}$-symmetric and broken regimes. Let us consider the Swanson Hamiltonian \cite{swanson2004transition}
    \begin{eqnarray}\label{Swanson Hamiltonian}
        \Tilde{H} = \frac{\omega_-}{2} p^2 + \frac{\omega_+}{2}x^2 + \frac{i}{2}\mathcal{A} \{x,p\} ,\quad i\hbar \partial_t \Tilde{\psi} = \Tilde{H}\Tilde{\psi},
    \end{eqnarray}
    where $p = -i \hbar \partial_x$, $\omega_\pm := \omega \pm (\alpha + \beta)$ and $\mathcal{A} :=\alpha-\beta$. According to  \cite{musumbu2006choice}, the Hamiltonian (\ref{Swanson Hamiltonian}) can be mapped via a similarity transformation to a harmonic oscillator, which corresponds to the top-right corner of the commutative diagram in Fig. \ref{fig: The square}
    \begin{eqnarray}\label{Swanson e.g. Time-dependent Schrodinger equation}
        &\Tilde{\eta}_i \Tilde{H} \Tilde{\eta}^{-1}_i = A_i (\alpha,\beta) p^2 + B_i(\alpha,\beta) x^2 =: \Tilde{h}_i,\\
        & \Tilde{\eta}_i \Tilde{\psi} = \Tilde{\phi}_i, \quad i \hbar \partial_t \Tilde{\phi}_i = \Tilde{h}_i \Tilde{\phi}_i,
    \end{eqnarray}
    where the index $i$ labels the non-unique choices of the Dyson maps. The specific forms of the parameters $A_i (\omega,\alpha,\beta) $ and $B_i (\omega,\alpha,\beta) $  are fixed by assuming at least two operators to be mapped to their Hermitian counterparts \cite{scholtz1992quasi}. Below we list three examples taken from \cite{musumbu2006choice}
    \begin{widetext}
    \begin{eqnarray}
        A_1 &=& \frac{\omega - 2 \sqrt{\alpha \beta}}{2 \omega} , \quad B_1 = \frac{\omega\left(\omega+ 2 \sqrt{\alpha \beta}\right)}{2} : \Tilde{\eta}_i H \Tilde{\eta}_i^{-1} = \Tilde{h}_i ,\quad \Tilde{\eta}_i N \Tilde{\eta}^{-1}_i = N ,\label{Eq. Dyson map 1}\\
        A_2 &=& \frac{\omega - \alpha -\beta}{2 \omega}, \quad B_2 = \frac{\omega}{2}\frac{\omega^2 - 4 \alpha \beta}{\omega - \alpha -\beta }: \Tilde{\eta}_i H \Tilde{\eta}_i^{-1} = \Tilde{h}_i ,\quad \Tilde{\eta}_i x \Tilde{\eta}_i^{-1} = x ,\label{Eq. Dyson map 2}\\
        A_3 &=& \frac{\omega^2 - 4 \alpha \beta}{2 \omega \left(\omega +\alpha + \beta\right)} , \quad B_3 = \frac{\omega \left(\omega + \alpha +\beta \right)}{2 } : \Tilde{\eta}_i H \Tilde{\eta}_i^{-1} = \Tilde{h}_i ,\quad \Tilde{\eta}_i p \Tilde{\eta}^{-1}_i = p ,\label{Eq. Dyson map 3}
    \end{eqnarray}
    \end{widetext}
    where $N$ is a number operator. 

    The average energy (\ref{Eq. average energy}) of the Hamiltonian is computed to 
    \begin{eqnarray}
        \braket{E} = (n+1/2) \sqrt{\omega^2 - 4 \alpha \beta} = (n+1/2) \sqrt{\mathcal{A}^2 + \omega_+ \omega_-}\nonumber
    \end{eqnarray}
    for $n\in \mathbb{N}$. The $\mathcal{PT}$ symmetry of the Swanson model is broken when $\omega^2 - 4 \alpha \beta =\mathcal{A}^2 + \omega_+ \omega_-< 0 $. Therefore in the $\mathcal{PT}$-broken regime, the average energy becomes complex. This is a common feature of $\mathcal{PT}$-symmetry quantum mechanics. We will consider the time-dependent boundary to mend this complex energy analog to \cite{fring2017mending}.
    
    The Schr\"{o}dinger equation (\ref{Swanson e.g. Time-dependent Schrodinger equation}) can be transformed by the unitary map (\ref{Eq. Unitary map 1}) to give the time-dependent Hermitian Schr\"{o}dinger equation $i \hbar \partial_t \phi_i  =h \phi_i$ corresponding to the bottom right corner of the commutative diagram in Fig. \ref{fig: The square}. The explicit form of the time-dependent Hermitian Hamiltonian is 
    \begin{eqnarray}
        h_i(t,x):=    \frac{\ell_t}{2\ell}\left\{x, i \hbar\partial_x\right\} - \frac{\hbar^2  A_i }{\ell^2}\partial_x^2  + B_i \ell^2 x^2 .
    \end{eqnarray}
    for $i = 1,2,3$. The corresponding Schr\"{o}dinger equation is simplified by performing a further unitary transformation of the form $\phi_j  = c_1 \text{exp}\left({i \frac{\ell \ell_t}{4 A_j\hbar}x^2}\right)\varphi_j (t,x) $, which reduces the equation to 
    \begin{eqnarray}\label{Eq. effective Hamiltonian 1}
        0&=& i 4 \hbar A_j \ell^2 (\varphi_j)_t +\hbar^2 4 A_j^2 (\varphi_j)_{yy} \nonumber\\
        &&- \ell^3 (4 A_j B_j \ell+\ell_{tt})y^2 \varphi_j  ,
    \end{eqnarray}
    with $c_j$ denoting the normalization constant. It is useful to notice that the combination of two parameters $4A_j B_j = \omega^2 - 4 \alpha \beta =:\Omega$ takes the same form for all three examples (\ref{Eq. Dyson map 1}) - (\ref{Eq. Dyson map 3}). 
    
    The above equation can be reduced further into the effective Harmonic oscillator if we consider the solution to the equation $\ell^3 (\Omega \ell + \ell_{tt})=\kappa^2$ where $\kappa$ is some constant. This is an Ermakov-Pinney equation \cite{pinney1950nonlinear,ermakov2008second2}, which can be solved exactly. One of the solutions is 
    \begin{eqnarray}\label{Eq. Ermakov-Pinney solution}
        \ell^2_j (t) &=& \frac{\kappa}{A_j B_j} \sin^2 \left(2 \sqrt{A_j B_j}t\right) + \kappa \cos^2 \left(2 \sqrt{A_j B_j}t\right)\nonumber\\
        &&+ \frac{2(\kappa - 1/4)}{\sqrt{A_j B_j}}\sin\left(2 \sqrt{A_j B_j}t\right)  \cos \left(2 \sqrt{A_j B_j}t\right). \nonumber\\
    \end{eqnarray} 
    Introducing the new time variable $\tau = \int^t 1/ \ell^2$, we find the effective Harmonic oscillator
    \begin{eqnarray}\label{Eq. Hermitian Hamiltonian DU approach}
        i 4 \hbar A_j (\varphi_j)_\tau + \hbar^2 4 A^2_j (\varphi_j)_{yy} -\kappa^2 y^2 \varphi_j =0.
    \end{eqnarray}
    Let us consider the Ansatz $\varphi_j^n = \exp\left(-i \epsilon^n_j \tau_j / A_j \hbar\right)\chi_j^n (y)$. Then the above effective harmonic oscillator is reduced to a Sturm-Liouville eigenvalue problem
    \begin{eqnarray}\label{Eq. Effective Hamiltonian}
        -\partial_{yy}\chi_j^n + \frac{\kappa^2}{4 \hbar^2 A_j^2}y^2 \chi_j^n = \epsilon_j^n \chi_j^n.
    \end{eqnarray}
    where there exist odd and even solutions are given in terms of hypergeometric functions
    \footnotesize
    \begin{eqnarray}
        {\chi_{\text{odd}}}_j^n (y) &=& e^{-\frac{1}{2}\frac{\kappa y^2}{2 \hbar A_j }} \left(\frac{\sqrt{\kappa}y}{\sqrt{2\hbar A_j}}\right){_1}F_1 \left[\frac{3}{4} - \frac{1}{4} \frac{2 \hbar A_j}{\kappa} \epsilon_j^n , \frac{3}{2} , \frac{\kappa y^2}{2 \hbar A_j }\right],\nonumber \\
        {\chi_{\text{even}}}_j^n (y)&=& e^{\frac{1}{2}\frac{\kappa y^2}{2 \hbar A_j }}  {_1}F_1 \left[\frac{1}{4} + \frac{1}{4} \frac{2 \hbar A_j}{\kappa} \epsilon_j^n , \frac{1}{2} , \frac{-\kappa y^2}{2 \hbar A_j }\right].\nonumber
    \end{eqnarray}
    \normalsize
    Therefore we found the solution to the effective Schr\"{o}dinger equation corresponding to the bottom right corner of the commutative diagram shown in Fig. \ref{fig: The square}
    \begin{eqnarray}\label{Eq. Metric-unitary wave function}
        \phi_j^n (t,x) = c^j_n e^{i \frac{\ell \ell_t}{4 A_j \hbar}x^2 - i \frac{1}{A_j \hbar}\epsilon_j^n \tau_j}\chi_j^n (x).
    \end{eqnarray}
    where the constants $c_n^j$ are fixed by normalisation $ 1=\braket{\phi_j^n | \phi_j^n} \implies c_n^{-2} = \int_{-1}^{1} dy \chi_n^\dagger \chi_n  $.

    The solution (\ref{Eq. Metric-unitary wave function}) can be mapped back to $\tilde{\phi}$ by use of an inverse mapping with the unitary transformation $u^\dagger \phi(t,x) = \phi(t, x / \ell(t)) / \sqrt{\ell(t)}$, which gives 
    \begin{eqnarray}\label{Eq. Metric-unitary wave function tilde}
        \tilde{\phi}_j^n (t,x) = \frac{c_n^j}{\sqrt{\ell(t)}} e^{i \frac{\ell \ell_t}{4 A_j\hbar}\left(\frac{x}{\ell(t)}\right)^2  - i \frac{1}{A_j \hbar}\epsilon_j^n \tau_j}\chi_j^n \left(x / \ell(t)\right).
    \end{eqnarray}
    Using this solution together with the Hamiltonian (\ref{Swanson e.g. Time-dependent Schrodinger equation}), one can calculate the average energy (\ref{Eq. Average energy 1}).
    \subsection{Average energy}
        {The quantum Fermi accelerator commonly refers to the quantum harmonic oscillator with a time-dependent boundary condition. It was first introduced in \cite{jose1986study} and one of its characteristics is its infinite increase of the average energy over time \cite{seba90}. It was later shown that with some specific oscillation of the boundary condition \cite{makowski1991}, the average energy shows periodic gain and loss but zero net increase. We will show in this section that in the $\mathcal{PT}$-symmetric case, the behavior of the average energy coincides with the result of \cite{makowski1991}, and in the $\mathcal{PT}$-broken case, we observe a new behavior of the average energy where the periodicity of the average is lost.}
            
        Let us plot the average energy (\ref{Eq. Average energy 1}) for three different Dyson maps (\ref{Eq. Dyson map 1}) - (\ref{Eq. Dyson map 3}) in the $\mathcal{PT}$-symmetric ($\Omega >0$) and the $\mathcal{PT}$-broken ($\Omega<0$) regimes. 
        \begin{figure*}[t]
        \centering
        \begin{minipage}[b]{0.9\textwidth}           \includegraphics[width=\textwidth]{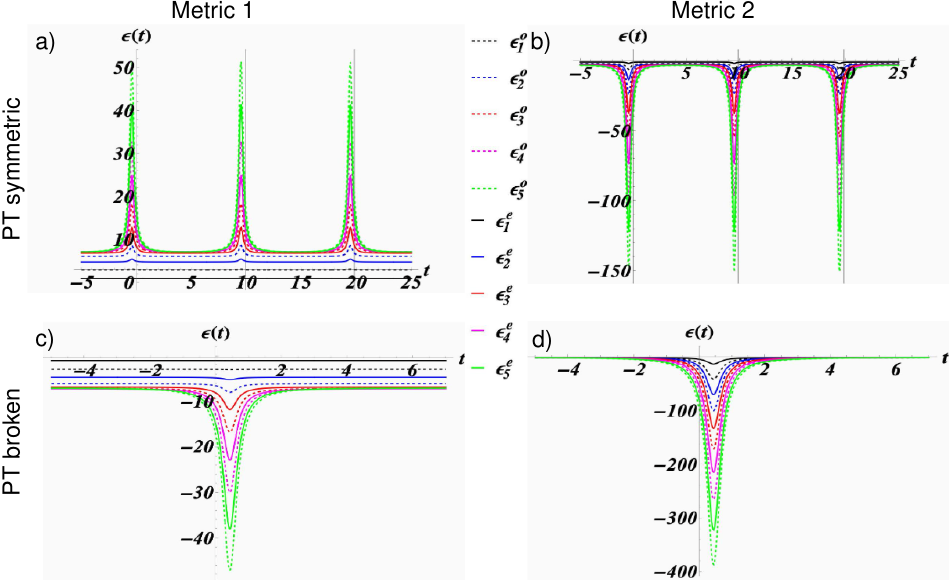}
            \end{minipage}
            \caption{Average energy for the $\mathcal{PT}$-symmetric/broken regimes over time for the first metric (\ref{Eq. Dyson map 1}) in panels (a) and (c), respectively. The results computed with the second metric (\ref{Eq. Dyson map 2}) for $\mathcal{PT}$-symmetric/broken regimes are shown in panels (b) and (d), respectively. The case involving the third metric (\ref{Eq. Dyson map 3}) is omitted as it is almost identical to the first metric with a slight scale difference.}
            \label{fig: PT Metric-Unitary energy spectrum}
        \end{figure*}

        In the $\mathcal{PT}$-symmetric regime shown in  Fig. \ref{fig: PT Metric-Unitary energy spectrum} panel (a) and (b), the average energy exhibits the periodic structure with $T = n\pi /2 \sqrt{A_j B_j}$, $n\in \mathbb{Z}$ for all three metrics (\ref{Eq. Dyson map 1})-(\ref{Eq. Dyson map 3}). This is because the average energy's periodicity is inherited from the boundary function (\ref{Eq. Ermakov-Pinney solution}), where the combination $A_i B_i$ is equal for all metrics. This finding leads us to the same conclusion as in \cite{makowski1991exactly}, indicating that although the average energy experiences time-dependent fluctuations, it remains periodic with no net gain or loss over a long time.

        In the $\mathcal{PT}$-broken regime, the real-valued average energy is consistent with previous observations \cite{fring2017mending}. In Figs. \ref{fig: PT Metric-Unitary energy spectrum} (c) and (d), {we observe a new behavior of the quantum Fermi accelerator, where} the average energy loses its periodic structure in this regime due to the non-periodic behavior of the boundary function (\ref{Eq. Ermakov-Pinney solution}), which hyperbolically diverges with time. Despite the divergent nature of the boundary function, the average energy remains constant over time and only experiences gain and loss near the origin. 
        
\begin{figure}[t]
        \centering
        \begin{minipage}[b]{0.5\textwidth}           \includegraphics[width=\textwidth]{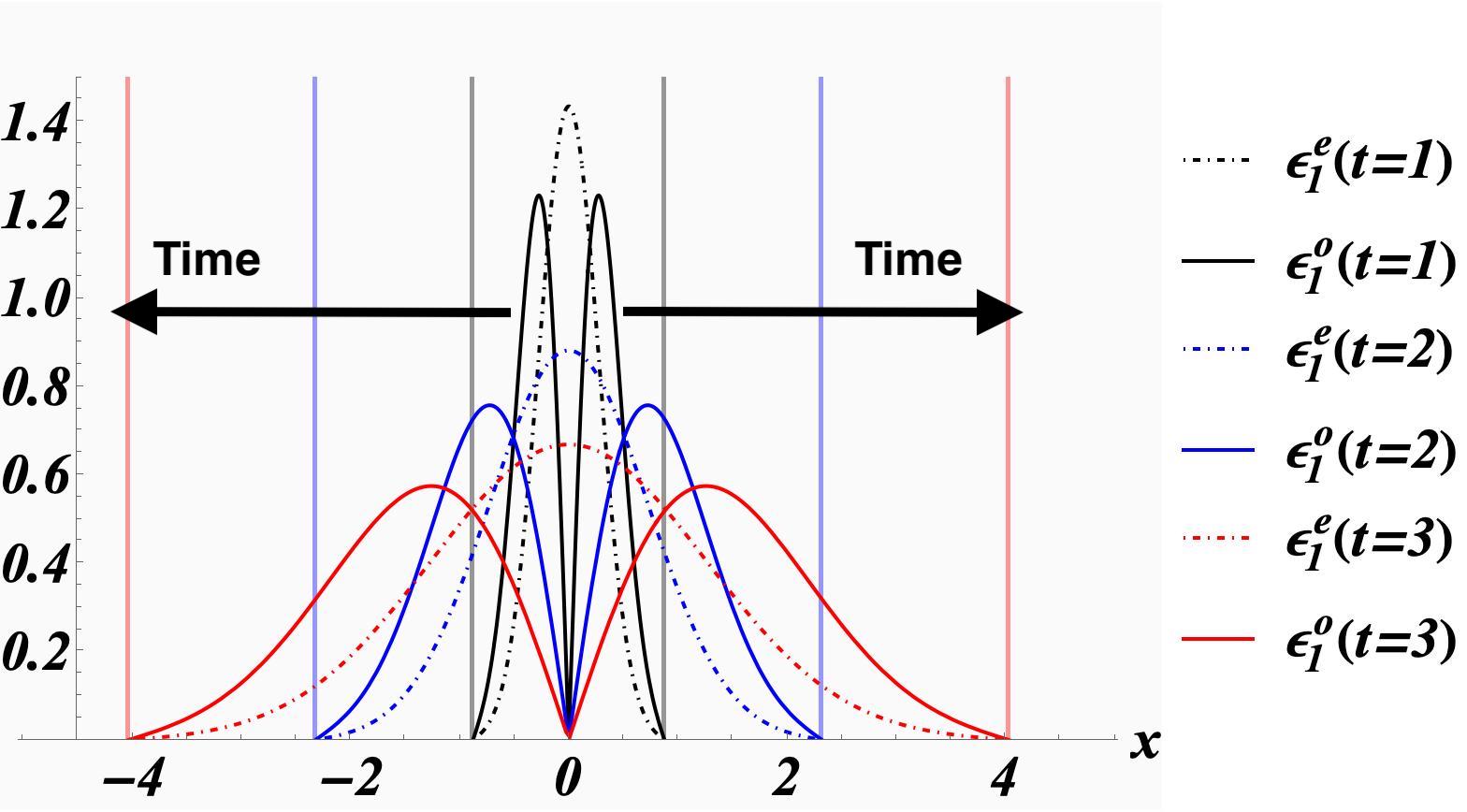}
            \end{minipage}
            \caption{Showing the infinite spreading of the probability density $\Tilde{\phi}^\dagger \Tilde{\phi}$ of the wave function (\ref{Eq. Metric-unitary wave function tilde}) with time in $\mathcal{PT}$-broken regime. The vertical line is the boundary $\ell(t)$, which is found by solving the Ermakov-Pinney equation. In the $\mathcal{PT}$-symmetric regime, the boundary moves periodically with a similar spreading of the probability density.}
            \label{fig: Spreading}
        \end{figure}

        {Furthermore, we observe a novel behavior of the non-Hermitian system where the probability density is infinitely spreading as the boundary moves away, which ensures the conservation of the probability even in the $\mathcal{PT}$-broken regime as demonstrated in Fig. \ref{fig: Spreading}.} This behavior is similar to that observed in single-particle open quantum systems \cite{hatano2014time}, but it differs from the context considered here in time-dependent pseudo-Hermitian non-Hermitian systems, where the non-Hermitian term does not result from environmental effects, as in \cite{hatano2014time}.

\section{Swanson model: Equivalence of time-dependent boundary and Dyson map}
    This section illustrates the commutativity of the diagram shown in Fig. \ref{fig: The square}. 

    Let us begin with the Swanson model (\ref{Swanson Hamiltonian}). Performing the unitary transformation (\ref{Eq. Unitary map 1}), the time-dependent non-Hermitian Hamiltonian is given in (\ref{Eq: Time-dependent non-Hermitian Hamiltonian}), where its explicit form is found to 
    \begin{eqnarray}\label{Eq: rewritten schrodinger eq}
        i \hbar\partial_t \psi&=& -\frac{\omega_- \hbar^2}{2\ell^2}\psi_{yy}+\frac{\omega_+ \ell^2}{2}y^2\psi  \nonumber\\
        &&+ {\hbar \mathcal{A}y \psi_y}  +  \frac{\hbar \mathcal{A}}{2} \psi +\frac{\ell_t }{2\ell} \left\{y,i \hbar \partial_x\right\}\psi
    \end{eqnarray}
    Similar to the previous section, one can perform further unitary transformations by $\psi = \exp (i \ell \partial_t \ell / 2 \hbar \omega_- y^2) \varphi (t,y) $ to the above equation. Let us consider the following Dyson map
    \begin{eqnarray}
        \eta &=& e^{- \frac{1}{2 \hbar \omega_-} \mathcal{A} \ell^2 y^2}\\
        \eta \psi &=& \eta e^{ \frac{1}{2 \hbar \omega_-} i LL_t y^2 } \psi  (t,y) \nonumber\\
        &=& e^{ \frac{1}{2 \hbar \omega_-} \left(-\mathcal{A} L^2 + i LL_t\right)y^2  }\varphi  (t,y) ,
    \end{eqnarray}
    which maps the non-Hermitian Hamiltonian to Hermitian Hamiltonian 
    \begin{eqnarray}\label{eq. first form of reduced Schrodinger}
    i 2 \hbar \omega_- \ell^2 \varphi_t + \hbar^2\omega_-^2 \phi_{yy} - \ell^3 \left(\Omega \ell + \ell_{tt}\right) y^2 \phi =0. 
    \end{eqnarray}
    Rescaling the variable as $y = \sqrt{\omega_- / 2 A_i} z$, the above equation is mapped to the effective Hamiltonian (\ref{Eq. effective Hamiltonian 1}), rendering the equivalence of two approaches.  
    
\section{Conclusion}
Our main finding is that time-dependent boundary conditions can be simulated with time-dependent metric operators and vice versa. In turn, this implies that the spontaneously broken $\mathcal{PT}$ regime can be mended, in the sense of acquiring real energies, not only by a time-dependent metric but equivalently also with time-dependent boundaries.
We demonstrated our assertions for the exactly solvable pseudo-Hermitian Swanson model. For this model, the time-dependent boundary functions are restricted by the Ermakov-Pinney equation. The characteristic behavior of this function, which is periodic in time or divergent, is inherited by the time-dependent average of the energy function. These restrictions may be relaxed at the cost of the model no longer exactly solvable.

In the $\mathcal{PT}$-symmetric regime, we find an oscillatory behavior of the average energy similar to the one found in \cite{makowski1991} for the harmonic oscillator with time-dependent coefficients. Different types of metric operators distinguish between whether this function has well-localized minima or maxima. In the spontaneously broken $\mathcal{PT}$-regime, the average energy is no longer periodic and develops only one well-localized minimum, irrespective of the choice of the metric.
\section*{Acknowledgment}

TT is supported
by JSPS KAKENHI Grant Number JP22J01230

\bibliographystyle{apsrev4-1}
\bibliography{apssamp}

\end{document}